\documentclass{iau}
\usepackage{graphicx}

\newcommand*{\Rm}{{\rm Rm}}
\newcommand*{\Ray}{{\rm Ra}}

\newcommand*{\Pm}{{\rm Pm}}
\newcommand*{\q}{{\rm q}}
\newcommand*{\Ek}{{\rm E}}
\newcommand*{\Ro}{{\rm Ro}}

\def\bfk{\mbox{\bf k}}
\def\bfg{\mbox{\bf g}}

\def\bfu{\mbox{\bf u}}
\def\bfB{\mbox{\bf B}}
\def\bfnabla{\mbox{\boldmath $\nabla$}}

\title[Planetary and Stellar Dynamos] %% give here short title %%
{Mechanisms of Planetary\\ and Stellar Dynamos}

\author[Emmanuel Dormy, Ludovic Petitdemange \& Martin Schrinner]   %% give here short author list %%
{Emmanuel Dormy, Ludovic Petitdemange \& Martin Schrinner}

\affiliation{CNRS, Equipe MAG (ENS-IPGP),\\ LRA, D\'epartement de Physique,\\ Ecole Normale
  Sup\'erieure, \\ 24 rue Lhomond, 75005 Paris, France \\ email: {\tt dormy@phys.ens.fr}}

\pubyear{2013}
\volume{294}  %% insert here IAU Symposium No.
\pagerange{119--126}
\jname{Solar and Astrophysical Dynamos and Magnetic Activity}
\editors{A.G. Kosovichev, E.M. de Gouveia Dal Pino, \& Y.Yan, eds.}
\begin{document}

\maketitle

\begin{abstract}
We review some of the recent progress on modeling planetary and stellar
dynamos. Particular attention is given to the dynamo mechanisms and
the resulting properties of the field. We present direct numerical simulations
using a simple Boussinesq model.
These  simulations are
interpreted using the classical mean-field formalism. We investigate the
transition from steady dipolar to multipolar dynamo waves solutions varying
different control parameters, and discuss the relevance to stellar magnetic
fields. We show that owing to the role of the strong zonal flow, this
transition is hysteretic. In  the presence of
stress-free boundary conditions, the bistability extends over a wide
range of parameters.
\end{abstract}

\firstsection % if your document starts with a section,
\section{Introduction}
Magnetic fields of low-mass stars and planets are thought to 
originate from self-excited dynamo action in their convective 
interiors. 
The accepted theory, known as dynamo theory, describes the
transfer from kinetic to magnetic energy as an instability process. Above a
given threshold electrical currents, and thus magnetic fields, are amplified
by a turbulent flow of a conducting fluid.

Observations of the magnetic fields produced by direct numerical
simulations (DNS) of dynamo action appear to fall
into two categories: fields dominated by large-scale dipoles (such as the
Earth and a fully convective star), and fields 
{with smaller-scale and 
non-axisymmetric structures} (such as the Sun). Two kinds of 
different temporal behaviour have also been identified: very irregular
polarity reversals (as in the Earth), and quasi-periodic reversals (as in
the Sun). Since the Earth and the Sun provide the largest database of
magnetic field observations, these objects have been well studied and
described in terms of alternative physical mechanisms: the geodynamo
involves a steady branch of the dynamo equations, with
fluctuations and possibly polarity reversals, whereas the solar dynamo
takes the form of a propagating dynamo wave. The signature of this wave at
the Sun's surface yields the well-known butterfly-diagram (Sunspots
preferentially emerge at a latitude that is decreasing with time during the
solar cycle).

\section{Bifurcation diagrams}
\label{Morin}
Let us start by considering the origin of the Earth's magnetic field, which
remains a challenging issue for physicists. We consider the
magnetohydrodynamic system of equations in a rotating spherical shell.
The problem can be described in its simpler form by a set of coupled
partial differential equations written in the classical Boussinesq limit
The governing equations can then be written in non-dimensional form
\begin{eqnarray}
{\rm E_\eta} \, 
\left[\partial _t \bfu +  (\bfu \cdot \bfnabla) \bfu \right]
&=& 
- \bfnabla \pi 
+ {\rm E} \, \Delta u
- 2 \bfk \times \bfu \nonumber\\
&-&{\Ray} \, T \, \bfg 
+ \left(\bfnabla \times \bfB \right) \times \bfB\, ,
\end{eqnarray}   
\begin{equation}
\partial _t \bfB = \bfnabla \times (\bfu \times \bfB) 
+ \Delta \bfB \, ,
\label{induction}
\end{equation} 
\begin{equation}
\partial_t T + (\bfu \cdot \bfnabla) T 
= \q \Delta T \, ,
\end{equation} 
\begin{equation}
\bfnabla \cdot \bfu =  \bfnabla \cdot \bfB = 0 \, ,
\end{equation} 
where
\begin{equation}
{\rm E}=\frac{\nu}{\Omega L^2}\, ; \ 
{\rm Pm}=\frac{\nu}{\eta}\, ; \ 
{\rm E_\eta}=\frac{\rm E}{\rm Pm}\, ; \ 
{\rm q}=\frac{\kappa}{\eta}\, ; \ 
{\rm Ra}=\frac{\alpha g \Delta T L}{\nu \Omega}\, . 
\end{equation}

The state of this system is
fully characterised by four independent controlling parameters. The Ekman
number $\Ek$, which can be interpreted as measuring the ratio of the
period of rotation (the length of
the day in the case of the Earth) to a typical viscous timescale, this number is extremely small in
the case of the Earth's core (the Earth, as the Sun, is in rapid rotation). The magnetic
Prandtl number $\Pm$, measuring the ratio of a typical ohmic timescale to 
viscous timescale, it is a characteristic of the fluid and is minute for
all liquid metals (including liquid iron in the Earth's core). The
Roberts number ${\rm q}$, also characterizing the fluid and which
compares a thermal timescale to the ohmic timescale, this number is
comparable with $\Pm$. Finally the Rayleigh number
$\Ray$, which measures a ratio of driving forces to forces slowing down the
motion. Its value is difficult to assess in a simple Boussinesq model. This
will be the most obvious controlling parameter, which needs to be varied to
investigate dynamo properties.

\begin{figure}
\centerline{\includegraphics[width=0.33\linewidth,clip=true]{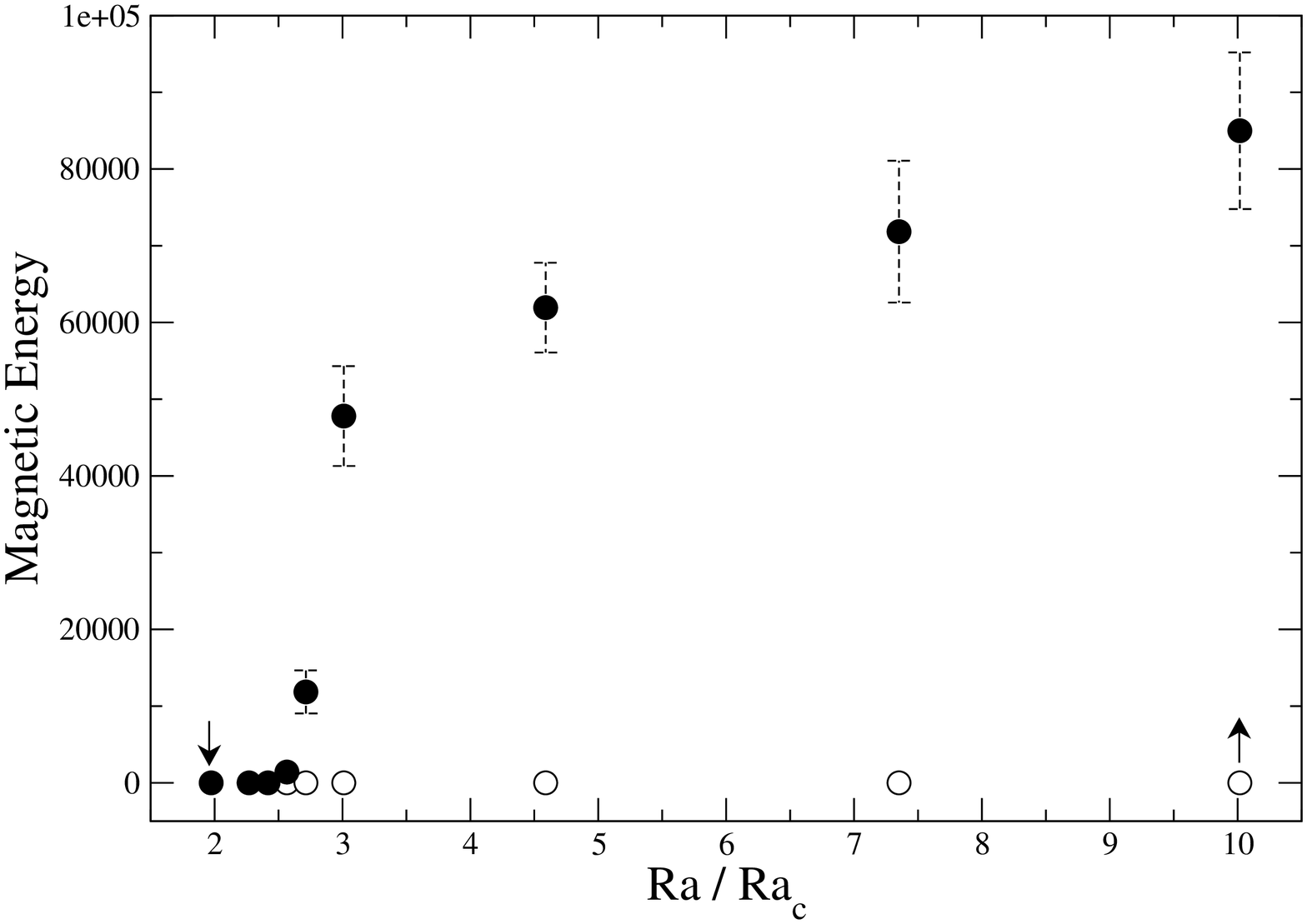}
\hskip -1mm\includegraphics[width=0.33\linewidth,clip=true]{FIGS/g_bifdyn_q3_E3e-4.eps}
\hskip -1mm\includegraphics[width=0.33\linewidth,clip=true]{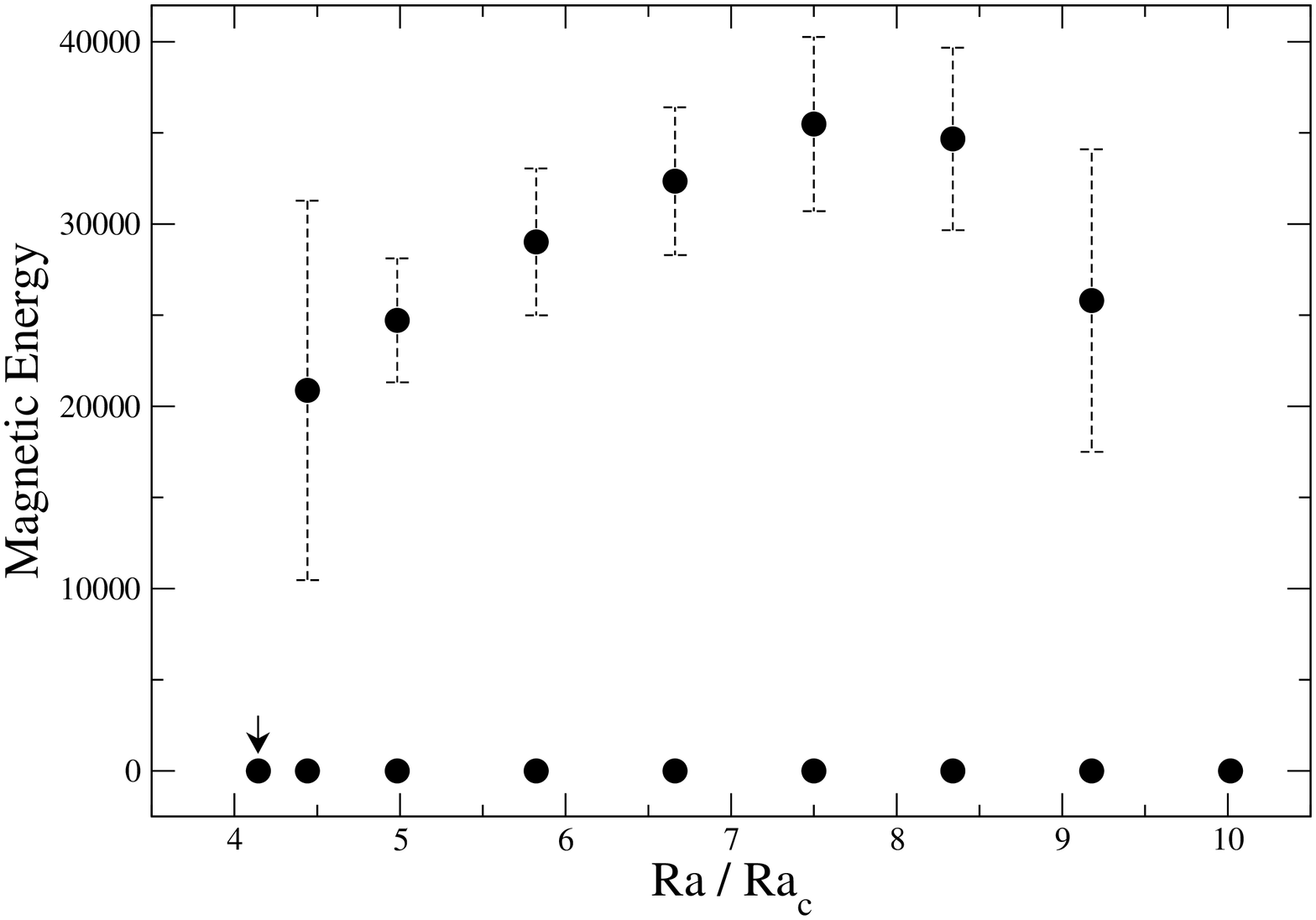}}
\caption{Mean magnetic energy as a function of the Rayleigh number. 
Error bars indicate the standard deviation of energy
fluctuations around the mean value.
All simulations are performed for $\Ek=3.10^{-4}$.
The bifurcation is found to be super-critical for $\Pm=6$ (left),
sub-critical for $\Pm=3$ (middle), and yields an isola diagram for $\Pm=1.5$
(right) [After
  \cite{MorinDormy}].}
\end{figure} 

We investigated in \cite{MorinDormy} the nature of the dynamo bifurcation in a configuration
applicable to the Earth's liquid outer core, i.e. in a rotating spherical shell with thermally
driven motions.
We show that the nature of the bifurcation, which can be either
super-critical or sub-critical or even take the form of isola (or detached
lobes) strongly depends on the parameters. This dependence is described in
a range of parameters numerically accessible (which unfortunately remains remote from
geophysical application), and we show how the magnetic Prandtl number
and the Ekman number control these transitions.

%% \begin{figure}
%% \begin{center}
%% \includegraphics[width=0.65\linewidth,clip=true]{FIGS/bifurcation_mag.eps}
%% \caption{Summary of the typical bifurcations obtained numerically. As the
%%   magnetic Prandtl number is decreased at fixed Ekman number one obtains
%%   successively supercritical (top), subcritical (middle) with
%%   restabilisation, and isola (bottom). The same sequence is achieved by
%%   increasing the Ekman number at fixed Rossby number.
%%   The dashed lines indicating unstable branches are here speculative,
%%   except for the ${\bf B}={\bf 0}$ state [After
%%   \cite{MorinDormy}].}
%% \label{bif_mag}
%% \end{center}
%% \end{figure}

We have studied in \cite{MorinDormy} different bifurcations obtained for Ekman number values ranging
from $10^{-3}$ to $10^{-4}$, magnetic Prandtl number values from $0.67$ to $6$ and
Rayleigh number values from $\Ray\simeq 2\times \Ray_c$ to $\Ray\simeq10\times
\Ray_c$.  In this parameter regime, for a given Ekman number, a super-critical
bifurcation  is obtained for a sufficiently high value of the
magnetic Prandtl number. By decreasing its value, it is possible to obtain a
sub-critical bifurcation, which may exhibit unusual features,
such as re-stabilization of the purely hydrodynamical state.
An unstable branch therefore must exist, it could be connected, for larger values of the Rayleigh
number, to the stable branch corresponding to dynamo solutions. If the
magnetic Prandtl number is further decreased, the range of Rayleigh number for which
the non dynamo solution is unstable vanishes. An isola is then obtained, in
this situation the purely hydrodynamical solution is 
always stable. The very same sequence is obtained by increasing $\Ek$ at
fixed $\Pm$, as the dipolar domain  shifts towards
lower values of $\Pm$ as $\Ek$ is decreased. We refer the reader to \cite{MorinDormy}
for further discussion on these aspects.

\section{From the Earth to the stars...}

\begin{figure}
\centerline{
\includegraphics[width=6cm,clip=true]{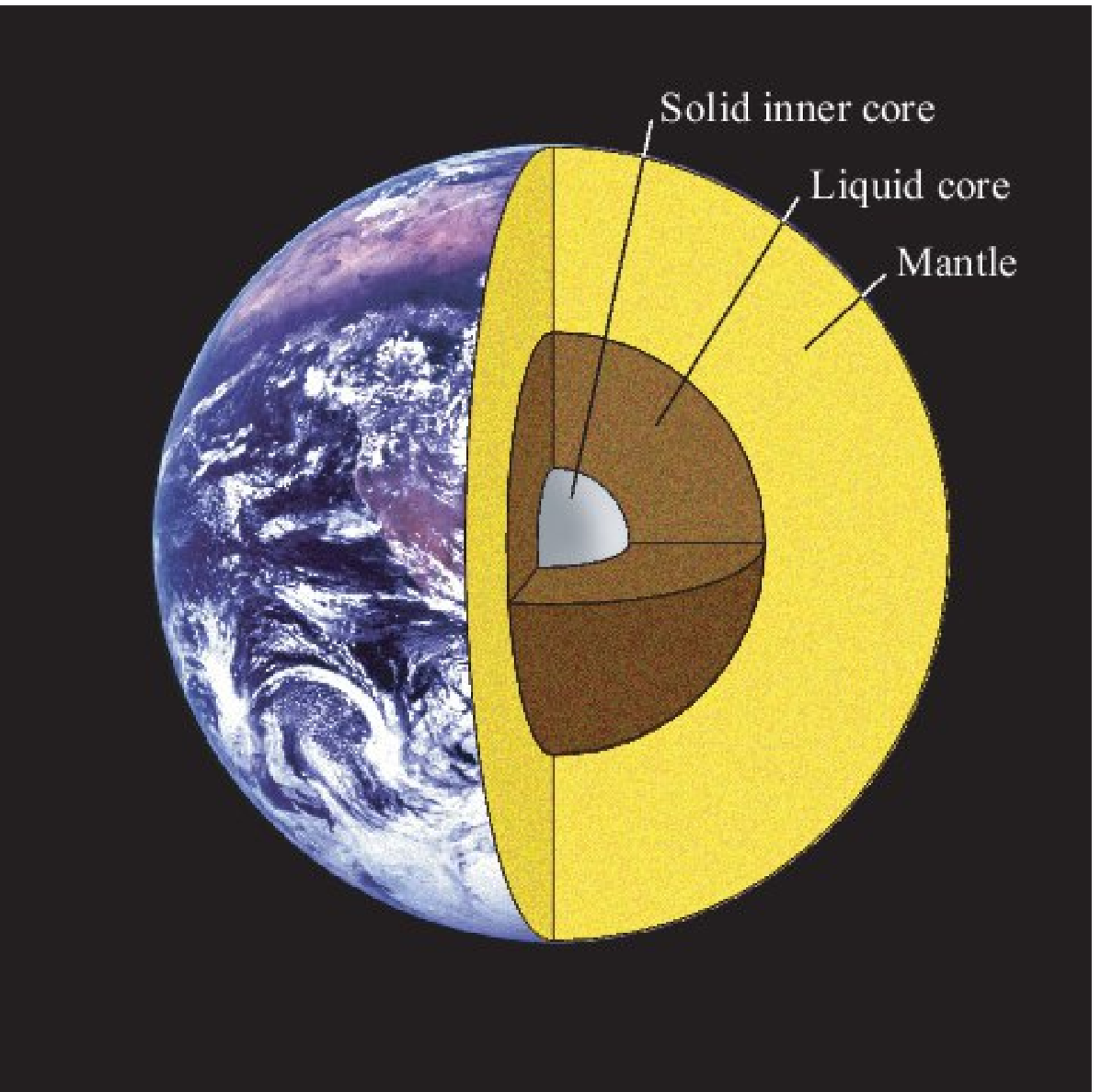}
\hskip 6mm
\includegraphics[width=6cm,clip=true]{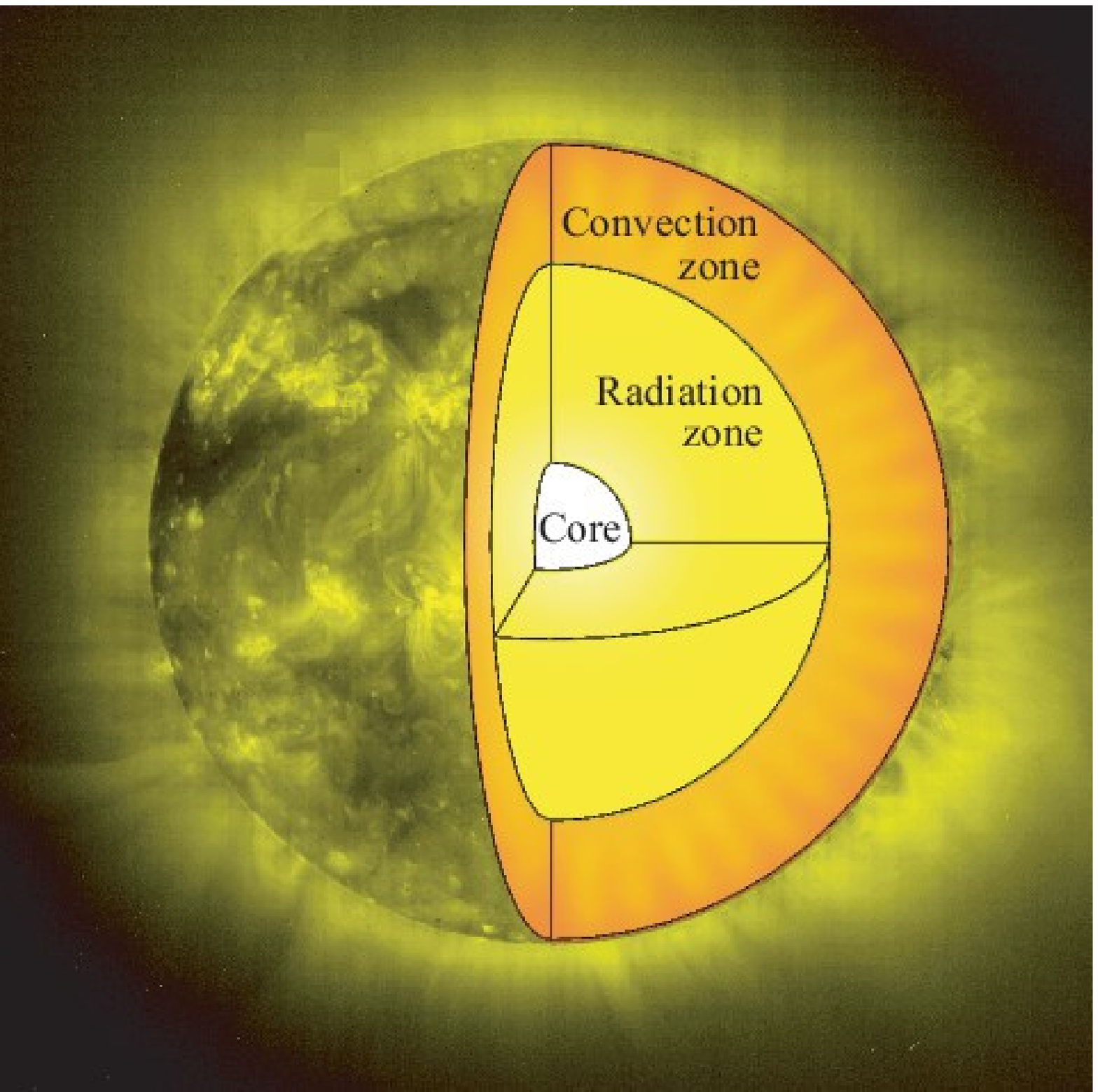}
}
\caption{Comparison of the Earth and Sun interior. The first striking
  difference regarding dynamo action is the aspect ratio of the dynamo
  region. The radiative zone of the Sun occupies a much larger fraction of
  the conducting region than the solid inner core of the conducting core 
[Figures from \cite{DS}].}
\label{cutaway}
\end{figure} 

Because of their very different natures (liquid metal in one case, plasma
in the other), planetary and stellar magnetic fields are studied by
different communities.
Non-dimensional numbers controlling the dynamics of the Earth and the Sun,
for example, do significantly differ (see~\cite[Zhang \& Schubert, 2006];
\cite[Tobias \& Weiss, 2007]{lelivre}).
As a practical matter however, the techniques 
{as well as the typical parameters}
used in numerical studies of these two systems are surprisingly similar. 
To some
extent this is due to the restricted parameter space available to present
day computations. 
{The parameter regime numerically accessible 
is rather remote from the actual objects. For planetary dynamos the 
main discrepancy relies in the rapid rotation in the momentum equation 
(characterized by the {\it Ekman number}), whilst 
for stellar dynamos it relies in solving the induction equation 
with weak resistive effects (characterized by high values of the 
{\it magnetic Reynolds number}).}

The relative success of numerical models to reproduce some of the key
characters to both geo (e.g. \cite[Glatzmaier \& Roberts, 1995]{BIB2})
and solar 
(e.g. \cite[Gilman, 1983]{BIB3} and \cite[Browning {\it et al}, 2006]{BIB4}) magnetic 
fields have lead us to argue that the aspect ratio of the dynamo region
(i.e. the radius ratio of the inner bounding sphere to the outer bounding
sphere) could be an essential parameter. Indeed, in the Earth, the inert
solid inner core extends to some 
35\% of the core radius, whereas in the Sun, the radiative zone fills
70\% of the solar radius (see Fig.\,\ref{cutaway}). 
One expects the convective zones of stars and
planets to have all possible intermediate aspect ratios, even extending
to fully convective spheres.
 
\begin{figure}[b]
% \vspace*{-2.0 cm}
\centerline{\includegraphics[width=15cm]{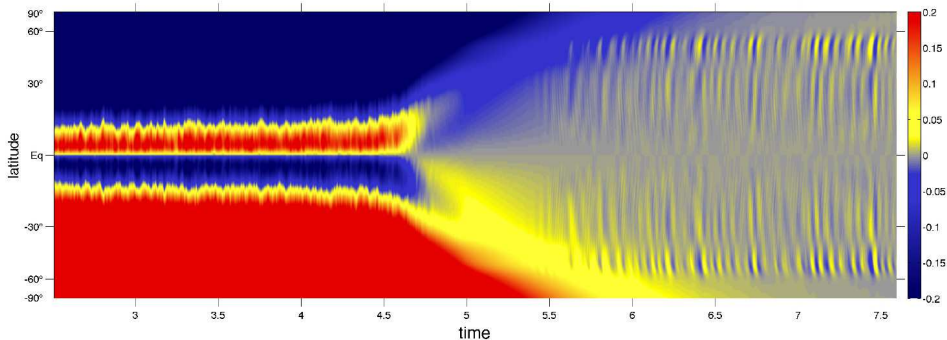}\ \  }
% \vspace*{-1.0 cm}
 \caption{Time evolution of the radial magnetic field averaged in longitude
(for an aspect ratio of $0.65$). The initial dipole field survives for a few
diffusion times, and then vanishes to yield a butterfly
diagram [After \cite{GoudardDormy}].}
   \label{GDfig1}
\end{figure}

In Goudard \& Dormy 2008, we showed that by varying the aspect ratio,
we could observe a sharp transition from 
a dipole dominated large scale-magnetic field to a cyclic dynamo with
a weaker dipole. Indeed in our simulations at fixed parameters, 
the strongly dipolar solution
becomes unstable with an increase of the aspect ratio above $0.65$. 
For this value --close to that of the Sun-- we observed that the strong dipole is first
maintained and then rapidly weakens, while dynamo action continues in a
different form: that of a wavy solution with quasi-periodic reversals
(Fig.\,\ref{GDfig1}), reminiscent of some aspects of the solar magnetic
field behavior.
This indicated that the geometry of the dynamo region could severely constrain
the existence of the dipole dominated solution.

\begin{figure}[b]
% \vspace*{-2.0 cm}
\begin{center}
 \includegraphics[width=4in]{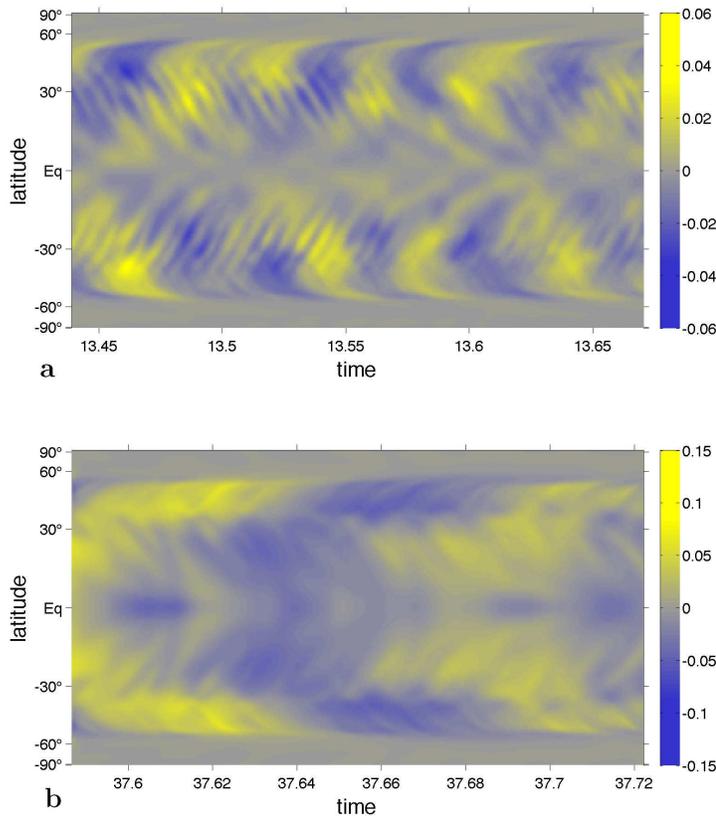} 
% \vspace*{-1.0 cm}
 \caption{Time evolution of
the zonal average of the azimuthal magnetic field below the surface of the
model, for an aspect ratio of $0.65$: the antisymmetric (a) and symmetric (b)
solutions [After \cite{GoudardDormy}].}
   \label{GDfig2}
\end{center}
\end{figure}

Because of the strong symmetry of the convective flows
influenced by the rapid rotation of the planet or the star,
it is well known that two independent families of solutions
can exist: with dipole symmetry (antisymmetric
with respect to the equator) and quadrupole symmetry
(symmetric with respect to the equator). We have indeed observed these two
families in our fully three dimensional simulations (Fig. \ref{GDfig2}).

\begin{figure}[b]
\begin{center}
 \includegraphics[width=4in]{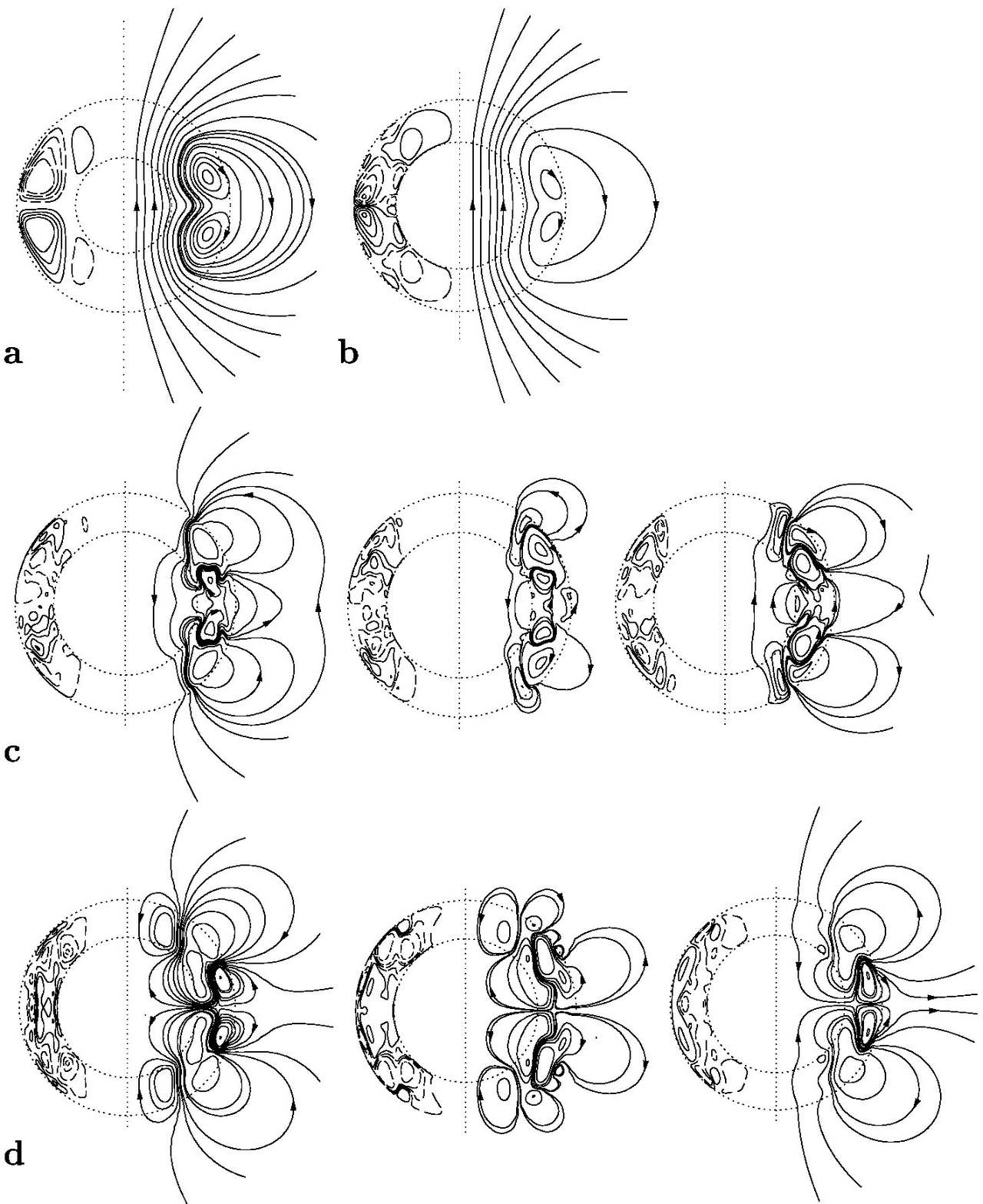} 
 \caption{The zonal average of the magnetic field in our 3D simulations.
Contours of the toroidal (east-west) part of the field are
plotted in the left hemisphere and lines of force of the meridional
(poloidal) part of the field plotted in the right hemisphere. 
The aspect
ratio is increased from $0.45$ (a) to $0.6$ (b) and to $0.65$ (c-d). The sequence
of dynamo waves is represented for the antisymmetric mode (c) and symmetric
mode (d). It is similar in nature to that produced by parametrized
models, see
\cite{Roberts72}  [After \cite{GoudardDormy}].}
   \label{GDfig3}
\end{center}
\end{figure}

Figure \ref{GDfig3} shows the azimuthally
averaged field for some of our fully 3D simulations. The
Earth-like mode is represented for aspect ratios of 0.45
and 0.6 (a and b). The active dynamo region lies outside
the tangent cylinder, it therefore gets increasingly
constrained as the inner sphere in increased. The dipole
eventually drops for large aspect ratio, when the volume
outside the tangent cylinder becomes too small.

Let us note that the above solutions 
(\cite[Goudard \& Dormy, 2008]{GoudardDormy}) 
have since then been reproduced using 
an independently written code 
by \cite{Simitevetal2010} and \cite{SimitevBusse2012}.

We shown that the steady dynamo branch can be replaced,
at larger aspect ratio, by an oscillatory dynamo mode.
Comparison with reduced parametrized models can help
interpret this transition to the solar-like mode. 
A strong zonal wind develops, in  the Solar-like regime.
This prompted us to suggest a transition from an $\alpha^2$ type to a dynamo of the $\alpha
\Omega$ type as the aspect ratio is increased. We shall see that this
requires a careful analysis.

\section{Dynamo Mechanisms}

Using the test-field method introduced in \cite{schrinner07,schrinner11}, 
we could compute mean-field dynamo coefficients
in \cite{AA}. These coefficients have been used in a mean-field 
calculation in order to explore the underlying dynamo mechanism.

\begin{figure}
  \centerline{\includegraphics[width=5cm]{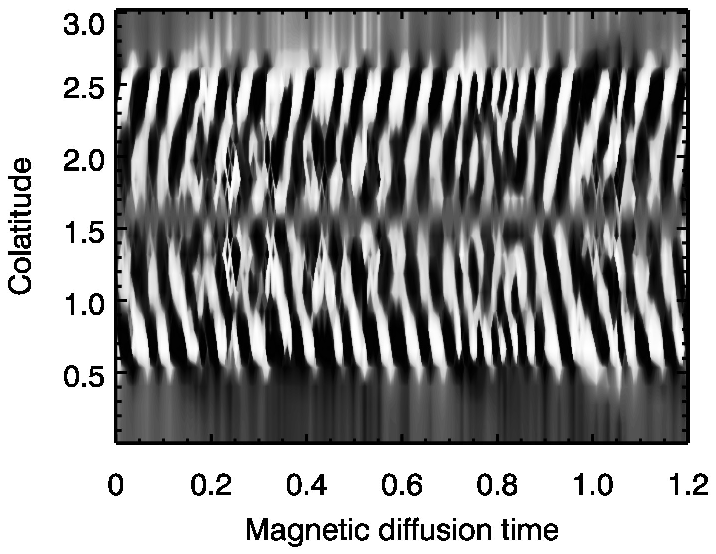}\hskip -6mm\includegraphics[width=5cm]{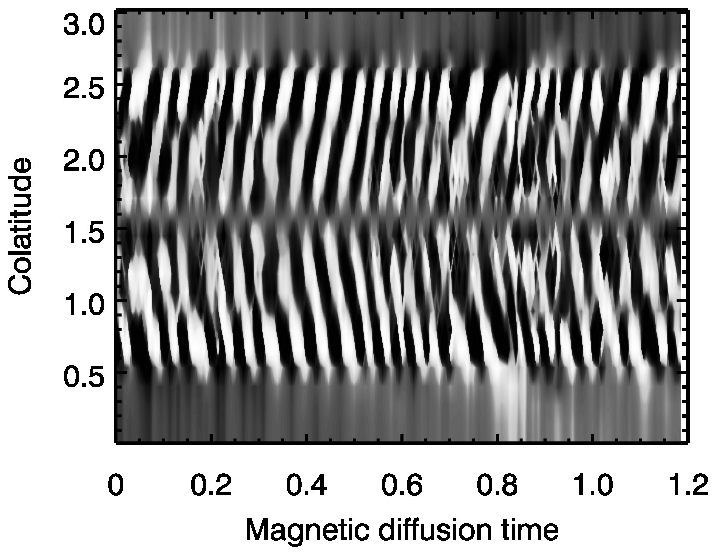}\hskip -6mm\includegraphics[width=5cm]{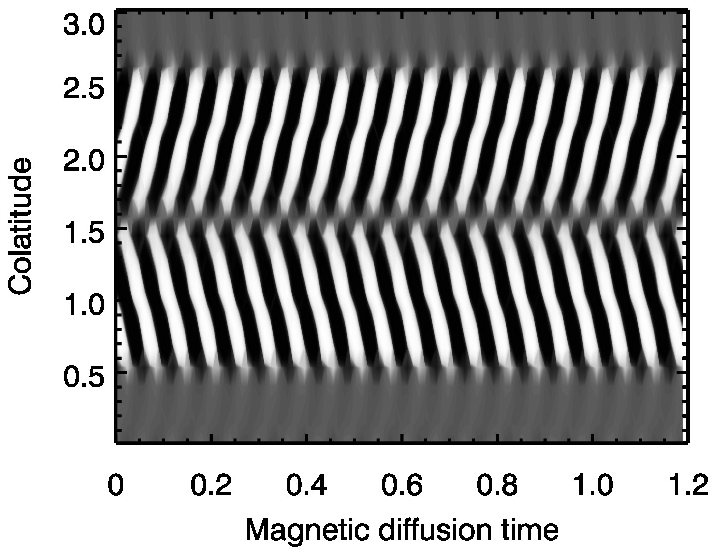}}
  \caption{Azimuthally averaged radial magnetic field at the outer shell 
  boundary varying with time (butterfly diagram) resulting from a 
  self-consistent calculation (left), kinematic calculation (middle) and 
  mean-field calculation (right). The contour plots have been normalised by 
  their maximum absolute value at each time step considered [After \cite{AA}].
  }
  \label{AAfig4}
\end{figure} 

The evolution of the magnetic field is cyclic. In Fig.~\ref{AAfig4}
(left), the azimuthally averaged radial magnetic field is represented at the outer 
shell boundary as a function of time (the so-called butterfly diagram).
A dynamo wave migrates away from the equator until it reaches 
mid-latitudes where the inner core tangent cylinder intersects the outer shell 
boundary. The magnetic field looks rather small scaled and multipolar.
This is confirmed by the magnetic energy spectrum which is essentially white, 
except for a negligible dipole contribution. Furthermore, the magnetic field is
weak, as expressed by an Elsasser number of 
\(\Lambda=B^2_{\mathrm{rms}}/(\mu\rho\eta\Omega)=0.13\). 

The kinematically advanced tracer field grows slowly in time, i.e. the model 
under consideration is kinematically unstable according to the classification
by \cite{schrinner10a}. But, deviations of the tracer field from the actual
field are hardly noticeable in the field morphology. Moreover, the very same
dynamo wave persists in the kinematic calculation (see also
\cite[Goudard \& Dormy, 2008]{GoudardDormy}), 
as visible in figure
\ref{AAfig4} (middle). Note that the tracer field in figure \ref{AAfig4} has 
evolved from random initial conditions. 
This time dependent mode offers a remarkable test of the mean-field
coefficient derived from the test-field approach.

A mean-field calculation based on the dynamo coefficients derived
using the test-field approach and the mean flow determined 
from the self-consistent model is presented in figure 
\ref{AAfig4}. The fastest growing eigenmodes form a conjugate complex pair and 
give rise to a dynamo wave which compares nicely with the direct numerical 
simulations. 

\begin{figure}
  \centerline{\includegraphics[width=5cm]{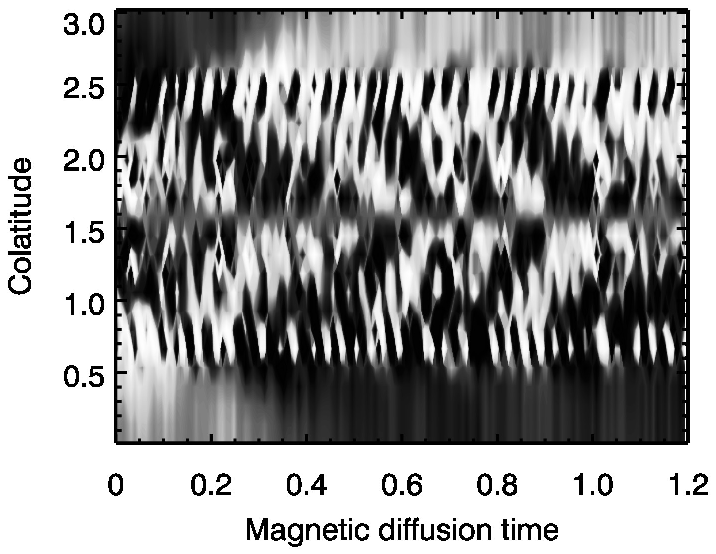}\hskip -2mm\includegraphics[width=5cm]{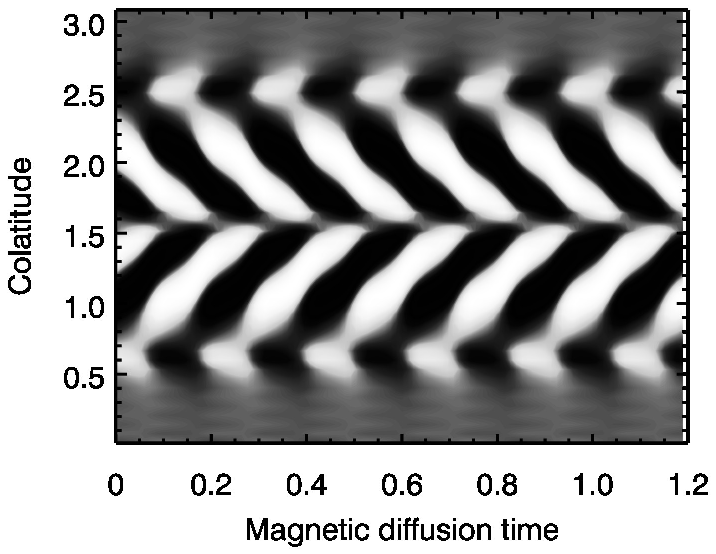}}
  \caption{Azimuthally averaged radial magnetic field at the outer shell 
  boundary varying with time (butterfly diagram) resulting from a 
  kinematic calculation with subtracted \(\Omega\)-effect (left) and 
  a corresponding mean-field calculation (right). The contour plots are 
  presented as in figure~\ref{AAfig4}  [After \cite{AA}].
  }
  \label{AAfig5}
\end{figure}

The influence of the differential rotation can be suppressed in the kinematic 
calculation of the tracer field without changing any other component of the flow. A 
butterfly diagram resulting from a kinematically advanced field is
presented in figure \ref{AAfig5} (left). Interestingly, the evolution 
of the magnetic field is again cyclic. 
The right chart of figure \ref{AAfig5} presents the butterfly diagram
corresponding to the fastest 
growing eigenmodes of the resulting \(\alpha^2\)-dynamo.

A particular dynamo mechanism does not seem to be responsible for the occurrence of 
periodically time-dependent magnetic fields. 
It turns out, that the influence of the large-scale radial 
shear (the \(\Omega\)-effect), is not necessary for cyclic field variations.
Instead, the action of small-scale convection
happens to be essential.
For the model presented here, small convective 
length scales are forced by a thin convection zone. 
Further investigations are needed to assess whether our finding is representative 
for a wider class of oscillatory models.

This study revealed that the oscillatory dynamo model under consideration is of 
the \(\alpha^2\Omega\)-type.
Although the rather strong differential rotation present in this model 
influences the magnetic field, the \(\Omega\)-effect alone is not 
responsible for its cyclic time variation. If the \(\Omega\)-effect is
suppressed the resulting \(\alpha^2\)-dynamo remains oscillatory. 
Surprisingly, the corresponding \(\alpha\Omega\)-dynamo leads to a 
non-oscillatory magnetic field.
The simpler assumption of an \(\alpha\Omega\)-mechanism therefore does not explain
 satisfactorily the occurrence of magnetic cycles.

\section{Dipole breakdown and bistability}
\label{ApJ}

We then investigated in \cite{ApJ} over seventy three-dimensional, self-consistent 
dynamo models obtained by direct numerical simulations. The control 
parameters, the aspect ratio and the mechanical boundary conditions 
have been systematically varied to build up this sample of models. 

\begin{figure}
  \centerline{\includegraphics[width=9cm]{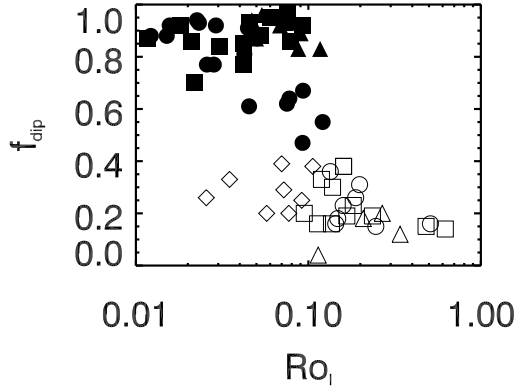}\ \hskip 1cm\ }
  \vskip -5mm
  \caption{Relative dipole field strength versus the local Rossby number 
  for all 72 models. Filled symbols stand for models dominated by a dipole field, open symbols 
  denote multipolar models. The symbol shape indicates different types of 
  mechanical boundary conditions: circles mean no-slip conditions at both 
  boundaries, triangles are models with a rigid inner and and a stress-free 
  outer boundary, and squares stand for models with stress-free
  conditions at both boundaries.
Simulations started from a weak magnetic 
field (diamonds) [After \cite{ApJ}].  
  }
  \label{fig_Rossby}
\end{figure} 

Both, strongly dipolar and multipolar models have been obtained. 
We could show in \cite{ApJ} that these dynamo 
regimes can in general be distinguished by the ratio of a typical convective 
length-scale to the Rossby radius (see \cite[Schrinner {\it et al}, 2012]{ApJ} for a precise
definition of $\Ro_\ell$). Models with a predominantly dipolar  
magnetic field were obtained, if the convective length scale is at least an 
order of magnitude larger than the Rossby radius (see Fig.~\ref{fig_Rossby}).

Moreover, we have highlighted
the role of the strong shear associated with the geostrophic zonal 
flow for models with stress-free boundary conditions. In this case the 
above transition disappears and is replaced by a region of bistability for 
which dipolar and multipolar dynamos co-exist (see again Fig.~\ref{fig_Rossby}).

There is a strong correlation between the topology and the time dependence of 
the magnetic field in dynamo models. Sudden polarity reversals or oscillations 
of the magnetic field do not occur in dipole dominated models in the low 
Rossby number regime. Conversely, reversals and oscillations are frequent in 
non-dipolar models with \(\Ro_\ell>0.1\) as well as in models with lower
local Rossby numbers with stress-free boundary conditions belonging to the 
multipolar branch. 

Whether non-dipolar models exhibit fairly coherent 
oscillations or irregular reversals of the magnetic field strongly depends on 
the magnetic Reynolds number. Coherent oscillatory solutions 
of the induction equation are most clearly visible in so-called butterfly 
diagrams; contours of the azimuthally averaged radial magnetic field at the 
outer boundary are plotted versus time and colatitude. 

\begin{figure}
  \centerline{\ \hskip7mm \includegraphics[width=10cm,height=4.1cm]{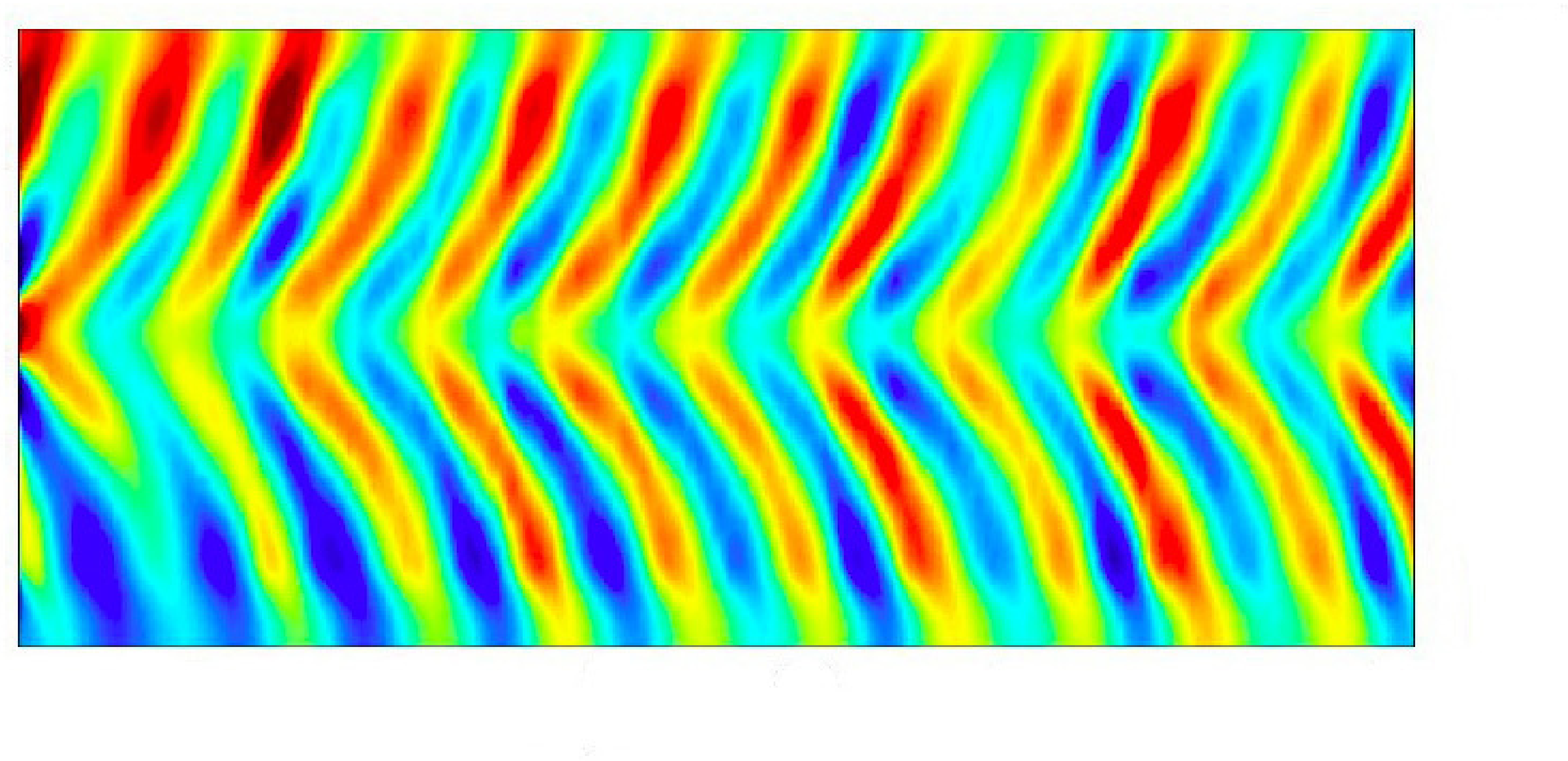}}
\vskip -4mm
  \centerline{\ \hskip7mm \includegraphics[width=10cm,height=4.1cm]{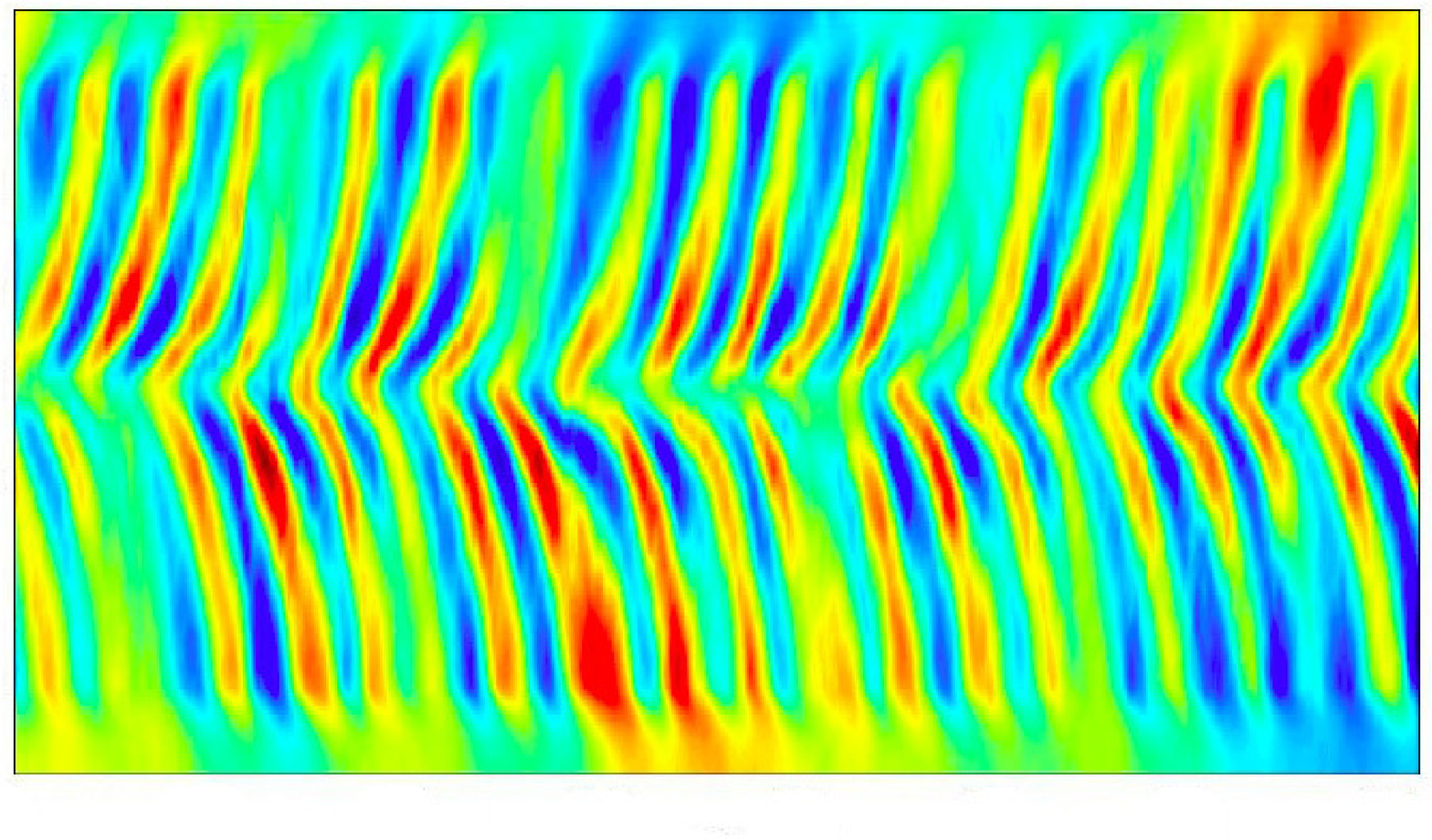}}
\vskip -4mm
  \centerline{\ \hskip7mm \includegraphics[width=10cm,height=4.1cm]{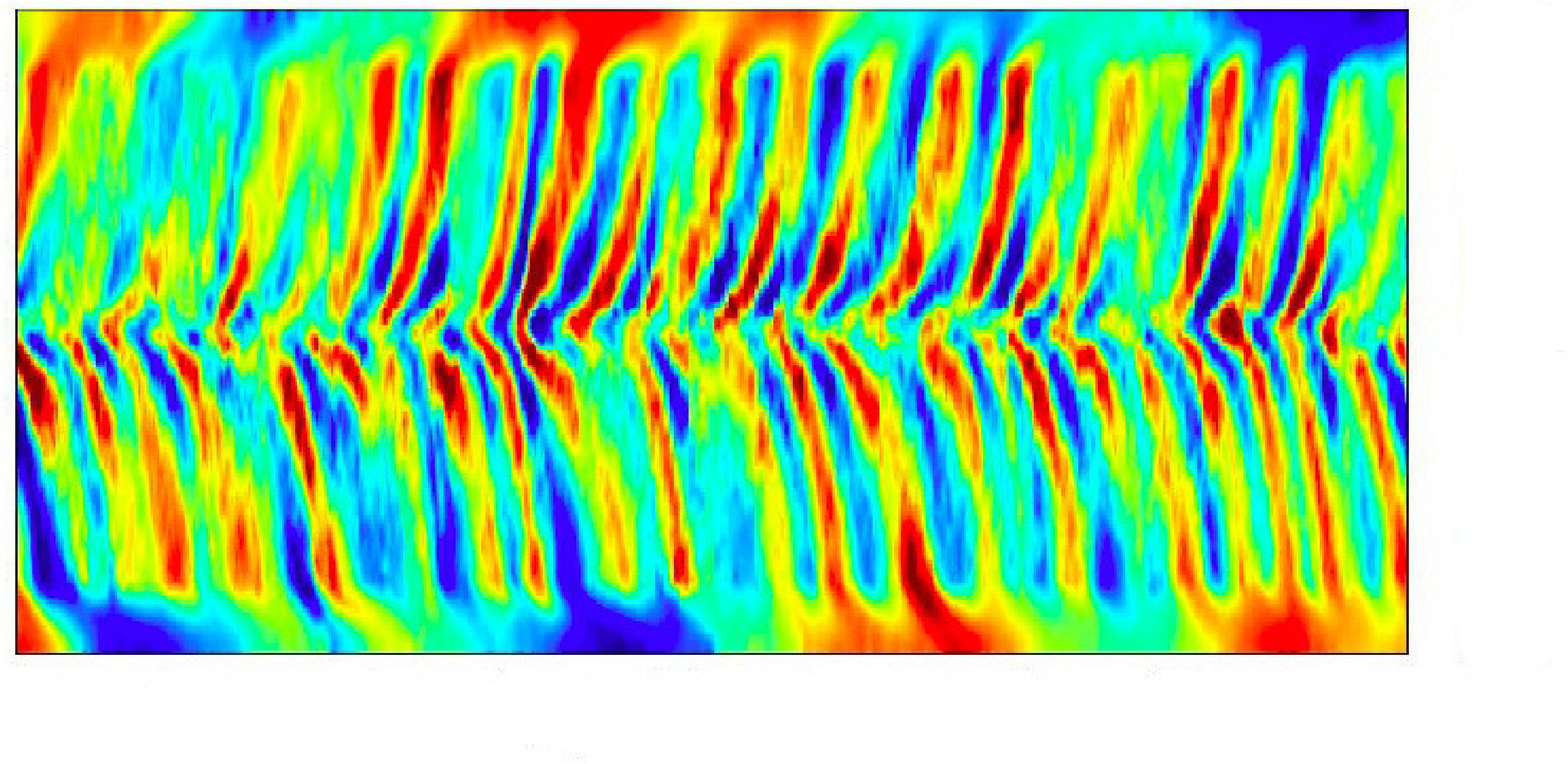}}
\vskip -4mm
  \centerline{\ \hskip7mm \includegraphics[width=10cm,height=4.1cm]{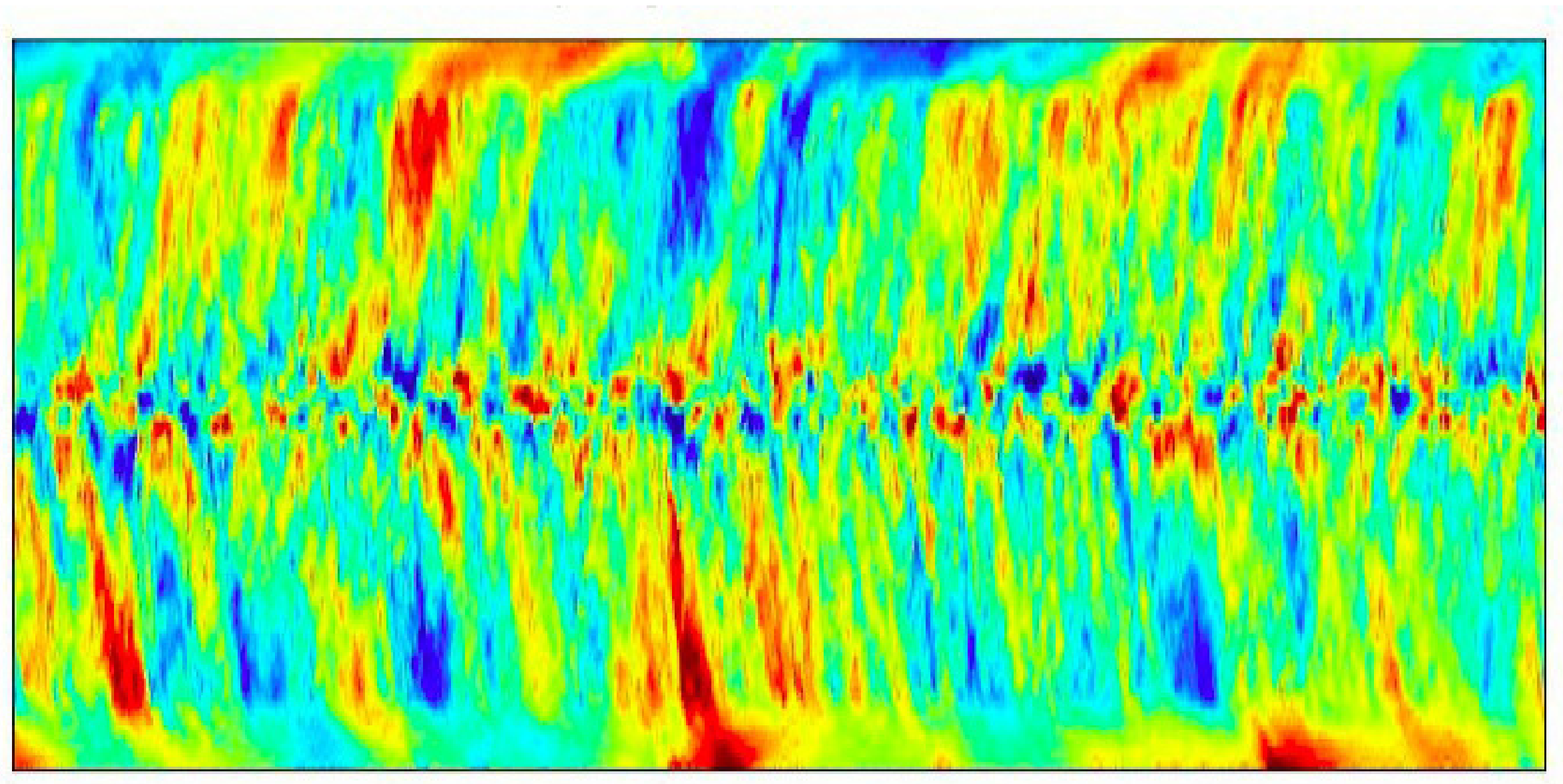}}
  \caption{The coherence of the butterfly diagram in DNS appears directly
    related to the magnetic Reynolds number $\Rm=\Ro \Pm / \Ek=U_{\rm rms}L/\eta$. Coherence is
    lost as the magnetic Reynolds number is increased. Here the
    diagrams respectively
    correspond (from top to bottom) to $\Rm = 75, \, 142, \, 270, \, 460$. }
  \label{fig_butt}
\end{figure}

Figure \ref{fig_butt} gives an illustration of this 
transition and at larger Rm is much less periodic and a cycle period cannot 
be identified. Dynamo models (in the non-dipolar regime) at higher magnetic 
Reynolds number exhibit even less temporal coherence. Following this somewhat 
arbitrary and qualitative criterion, we find that non-dipolar dynamos of our 
sample with \(\Rm\lesssim 200\) generate magnetic fields which vary 
periodically in time. The lower the magnetic Reynolds number, the more 
coherent is the time variability of the magnetic field. This has
important implications on the applicability of these models to stars
as the Sun with a well defined cycle.

\begin{figure}
\ \vskip 6mm
\centerline{\includegraphics[width=8cm]{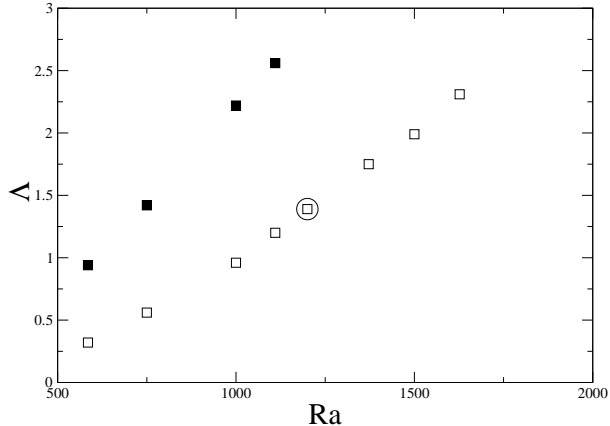}}
\caption{Evolution of the magnetic field strength, measured by the Elsasser
number, for both branches as the Rayleigh number is varied at fixed Ekman
and magnetic Prandtl numbers (E = $10^{-4}$ and Pm = $1$). Filled symbols stand for models
dominated by a dipole field, open symbols denote multipolar
models. These models used  stress-free conditions at both boundaries [After \cite{ApJ}].}
\label{Ra_Els_branch}
\end{figure} 

It is interesting to ponder on the transitions between the 
dipolar and multipolar branch for stress-free models when one 
single control parameter is varied. The two branches are illustrated 
in Fig. \ref{Ra_Els_branch} for a fixed Ekman number of \(E=10^{-4}\)
and magnetic Prandtl number of \(Pm=1\). For both branches, the local Rossby 
number increases with increasing Rayleigh numbers. If the 
Rayleigh number is increased from \(Ra=1110\)  on the dipolar 
branch to \(Ra=1200\), the relative dipole field strength collapses 
(the local Rossby number crosses the $\Ro_\ell\sim 0.1$ boundary).
The multipolar field configuration then appears to be the only stable 
solution (circle on the figure) and a hysteretic behavior is observed if the 
Rayleigh number is decreased from this state.

\section{Mean-field, Boussinesq and Anelastic: a hierarchy of models}

We have presented here several numerical developments relying on a simple
Boussinesq dynamo model. Assuming a constant reference density is clearly a
strong simplification of the governing equations. This approximation is
fairly reasonable for experimental setups, but harder to justify on the
large scale of planetary or stellar interiors. When the reference state is
stratified under its own weight, but the fluid velocity remains small
compared to the sound speed, it is safer to rely on an anelastic formalism
(different approximations can be designated by this name).

We are thus left with a hierarchy of tools in order to investigate
the magnetic field generation in planets and stars. 

The first approach,
which is also the lightest in terms of computational resources is the
mean-field formalism. This has originally been introduced as a modeling
tool. 
We have shown that it offered a remarkable tool to interpret the results of
DNS (based on the unparametrized partial differential equations).

The natural candidate to investigate stellar interiors is then to rely on
an anelastic formulation of the problem. This allows to to take into
account the variation of the reference density with depth.
Such approaches have been used for example by \cite{BIB4,Browning2010} in
the solar and stellar context.

The Boussinesq approach offers an intermediate description. It is free of
parametrisation, but lacks the description of effects indiced by stratification.
It is however striking that differential rotation profiles obtained with
this approach (e.g. \cite[Browning, 2010]{Browning2010}) are extremely similar to those
obtained using Boussinesq models.
In fact, since our results were published, \cite{Gastine} were able to
reproduce a similar set of transitions as those reported in
section~\ref{ApJ} but using an anelastic model. They observed that the
local Rossby number was the controlling parameter in anelastic models as well.

\section{Conclusion}

We have shown that information on stellar dynamics can be gained from direct
numerical models relying on a simple Boussinesq formulation. 
Boussinesq models  offer a simple and flexible tool to investigate
stellar dynamics and dynamo action in stars. As any model they require
careful interpretation and cannot account for stratification effects.
Boussinesq models provide
a useful tool to understand how the magnetic field behaviour is affected by
the controlling parameters.

Besides the bistability we reported, and which was since then also
observed in anelastic models, could be relevant to low mass stars.
Spectropolarimetric and spectroscopic observations of
two groups of very low mass fully-convective stars sharing similar stellar
parameters but generating radically different types of magnetic fields.
This bi-stability could be the equivalent, albeit involving inertial
rather than viscous effects, to the
weak-field versus strong-field bistability predicted for the geodynamo
(see \cite[Morin {\it et al}, 2011]{MorinDonati}).

\section*{Acknowledgments}
The ideas and results presented in this review were developed over
the last few years in collaboration with co-workers,
in particular:
Vincent Morin, Laure Goudard, Julien Morin and Jean-Fran\c{c}ois Donati.

\end{document}